\documentclass[reprint,superscriptaddress,amsmath,amssymb,aps,prb,floatfix]{revtex4-2} 
\usepackage{graphicx}
\usepackage{dcolumn}
\usepackage{bm}
\usepackage{physics}
\usepackage[colorlinks,
    linkcolor=blue,
    anchorcolor=red,
    citecolor=green
]{hyperref}
\usepackage{array}
\usepackage{cancel}
\usepackage{times}
\usepackage[caption = false]{subfig}
\usepackage{comment}

\newcommand{\mi}{\mathrm{i}}
\newcommand{\me}{\mathrm{e}}

\begin{document}

\title{Multichannel topological Kondo effect}

\author{Guangjie Li}
\affiliation{%
Department of Physics and Astronomy, Purdue University, West Lafayette, Indiana 47907, USA.
}%

\author{Yuval Oreg}
\affiliation{%
Department of Condensed Matter Physics, Weizmann Institute of Science,
Rehovot 76100, Israel
}%

\author{Jukka I. V{\"a}yrynen}%
\affiliation{%
Department of Physics and Astronomy, Purdue University, West Lafayette, Indiana 47907, USA.
} 
\affiliation{Purdue Quantum Science and Engineering Institute, Purdue University, West Lafayette, Indiana 47907, USA}

\date{\today}

\begin{abstract}
A Coulomb blockaded $M$-Majorana island coupled to normal metal leads realizes a novel type of Kondo effect where the effective impurity ``spin'' transforms under the orthogonal group $SO(M)$. 
The impurity spin stems from the non-local topological ground state degeneracy of the island and thus the effect is known as the topological Kondo effect. We introduce a physically motivated $N$-channel generalization of the topological Kondo model. 
Starting from the simplest case $N=2$, we conjecture a stable intermediate coupling fixed point and evaluate the resulting low-temperature impurity entropy. 
The impurity entropy  indicates that an emergent Fibonacci anyon can be realized in the $N=2$ model.
We also map the case $N=2$, $M=4$ 
to the conventional 4-channel Kondo model and find the  conductance at the intermediate fixed point. 
By using the  perturbative renormalization group, 
we also analyze the large-$N$ limit, where the fixed point moves to weak coupling. 
In the isotropic limit, we find an intermediate stable fixed point, which is stable to ``exchange" coupling  anisotropies, 
but unstable to channel anisotropy. 
We evaluate the fixed point impurity entropy and conductance to obtain experimentally observable signatures of our results. In the large-$N$ limit we evaluate the full cross over function describing the temperature-dependent conductance.
\end{abstract}

\maketitle

\textit{Introduction.} 
Since the original Kondo model of a magnetic impurity screened by a single orbital in a metal~\cite{kondo1964resistance}, its multichannel generalization~\cite{nozieres1980kondo}, especially in mesoscopic devices~\cite{Pustilnik_2004,Winston2021}, has proven to be a fruitful  system exemplifying exotic many-body phenomena~\cite{affleck1995conformal,1998AdPhy..47..599Z,2007Natur.446..167P,2015Natur.526..237K,2015Natur.526..233I,2018Sci...360.1315I} such as emergent anyonic excitations~\cite{PhysRevB.101.085141,PhysRevB.101.235131,PhysRevB.105.035151,Lotem22} even in the simplest two-channel Kondo (2CK) model~\cite{PhysRevB.46.10812,1994PhRvB..4910020S}. 
The key parameters that determine the behavior  of the system are the spin of the impurity~($S$) and the largest possible total spin~($N/2$) for $N$ channels of conduction electrons. 
When $N>2S$, namely the overscreened case~\cite{nozieres1980kondo}, the low-temperature fixed point is at intermediate coupling strength, with both weak and strong coupling fixed points unstable. 
Notably, in the limit of large~$N$, the intermediate coupling fixed point moves toward weak coupling and becomes perturbatively accessible in $1/N$ expansion~\cite{nozieres1980kondo,PhysRevLett.70.686}.

The intermediate coupling fixed point cannot generically be described by a Fermi liquid theory. 
For example, in a mesoscopic 2CK device, the conductance correction near $T=0$ is proportional to $T$~\cite{2007Natur.446..167P,2015Natur.526..233I}, and not to $T^2$ expected 
of a Fermi liquid~\cite{PhysRevLett.75.709}. 
Besides the conductance, the impurity entropy also shows exotic non-integer quantum dimension, which can be interpreted as a fractional ground state degeneracy. 
As shown by Emery and Kivelson~\cite{PhysRevB.46.10812} (see also Refs.~\cite{1994PhRvB..4910020S,Coleman_1995,Rozhkov_1998}), an emergent Majorana will 
remain of  the impurity spin 
after the screening by conduction electrons. The low-temperature impurity entropy is given by $\ln \sqrt{2}$ where $\sqrt{2}$ is the  quantum dimension of a single Majorana (two uncoupled Majoranas have a ground state degeneracy 2). 
For 3CK, the screened impurity entropy is $\ln\varphi$, where $\varphi=(1+\sqrt{5})/2$ is the Golden ratio \cite{doi:10.7566/JPSJ.90.024708,PhysRevLett.128.146803}, exhibiting the quantum dimension of a Fibonacci anyon. 
The conventional multichannel Kondo (MCK) models based on the $SU(2)$ symmetry group (which is natural in the case of a magnetic impurity) have been extensively studied by using conformal field theory
(CFT)~\cite{PhysRevLett.67.161,affleck1990current,affleck1991kondo,affleck1991critical,PhysRevB.48.7297,ludwig1994exact,affleck1995conformal,PhysRevLett.67.3160,PhysRevB.58.3794} and various other methods~\cite{PhysRevB.65.195101,PhysRevB.57.R5579,PhysRevB.54.14918,PhysRevLett.52.364,tsvelick1985exact,gan1994multichannel,PhysRevLett.70.686,gan1994multichannel,PhysRevLett.70.686,PhysRevB.80.205114,MARTINEK2007e343,PhysRevB.86.195128,PhysRevB.97.235450}.

It is natural to expect that the  rich physics of the multichannel Kondo effect can be further expanded by considering symmetry groups beyond the conventional $SU(2)$. 
Recently, B\'eri et al.~\cite{PhysRevLett.109.156803,PhysRevLett.110.216803} showed that a Coulomb blockaded topological superconductor hosting $M$ Majorana zero modes coupled to $M$ normal metal leads displays a Kondo interaction with $SO(M)$ symmetry~\footnote{See also Refs.~\cite{PhysRevLett.126.147702,PhysRevB.103.195131} for $SO(5)$ Kondo model without Majoranas. }. 
Even though this ``topological Kondo'' model has only one $SO(M)$ channel~\footnote{Since we consider normal metal leads, by 'channel' we mean a channel of itinerant complex fermions, equivalent to two  Majorana fermion  channels~\cite{tsvelik2014topological}.  }, 
in certain cases (such as $M=3,4$) it  can be mapped to an $SU(2)$ MCK model and therefore has non-Fermi liquid (NFL) behavior at low temperatures. 
For example, the conductance correction near $T=0$ is proportional to $T^{2(M-2)/M}$ and the impurity entropy generically indicates a fractional ground state degeneracy~\cite{Altland_2014}.  
With the $SO(M)$ symmetry group providing a relatively stable NFL fixed point, the  single-channel topological Kondo model has attracted a wide range of detailed studies and extensions~\cite{PhysRevB.94.235102,PhysRevB.96.205403,PhysRevB.99.014512,PhysRevResearch.2.043228,PhysRevB.97.235139,PhysRevB.89.045143,PhysRevLett.113.076404,PhysRevLett.114.116801,Buccheri_2016,PhysRevLett.119.027701,BUCCHERI201552,PhysRevB.104.205125}.  
However,  the multichannel version of it has not yet been studied.  
    
In this paper, we generalize the topological Kondo model to  $N \geq 2$ channels and propose a physical realization for it. 
Already in the relatively simple $N\!= \! 2$, $M \!= \! 8$ case we find a quantum dimension indicating an emergent Fibonacci anyon which cannot be realized for any $M$ in the single channel $SO(M)$ model. 
In this simplest 2-channel generalization, we also find a mapping between   $SO(4)$ and conventional 4CK model, allowing us to find the exact fractional fixed point conductance, Eq.~(\ref{eq:2CSO4}). 
In order to study a more general case, we  then introduce large-$N$ perturbation theory similar to what has been done with the $SU(2)$ case~\cite{kuramoto1998perturbative,gan1994multichannel, OregDGG2002}. 
We focus on the experimentally relevant observables of the conductance~\cite{2007Natur.446..167P,2015Natur.526..233I,PhysRevLett.90.136602} and impurity entropy~\cite{PhysRevLett.128.146803,hartman2018direct,PhysRevLett.123.147702,child2021entropy,pyurbeeva2022electronic,child2022robust}.

\textit{Two-channel topological Kondo model.}\label{sec:2channel}
The two-channel generalization of the isotropic topological Kondo interaction Hamiltonian is
\begin{align}
 H_K=\lambda_{1} \boldsymbol{S} \cdot \boldsymbol{J}_1(x_0)+\lambda_{2}\boldsymbol{S} \cdot \boldsymbol{J}_2(x_0), \label{eq:V2C}   
\end{align}
where $S^{(\alpha,\beta)}=-\mi \gamma_{\alpha} \gamma_{\beta}/2$ and $J_i^{(\alpha,\beta)}=-\mi( \psi^{\dagger}_{i,\alpha}\psi_{i,\beta}- \psi^{\dagger}_{i,\beta}\psi_{i,\alpha})$ are, respectively, the impurity and conduction electron ``spin'' operators 
that satisfy the $SO(M)$ algebra; the  vectors $\boldsymbol{S} $ and $\boldsymbol{J} $ are formed of $M(M-1)/2$ components labeled by $(\alpha,\beta)$ with $\alpha \neq \beta$ taking values from 1 to $M$. 

\begin{figure}[tb]
\centering
\includegraphics[width=0.98\columnwidth]{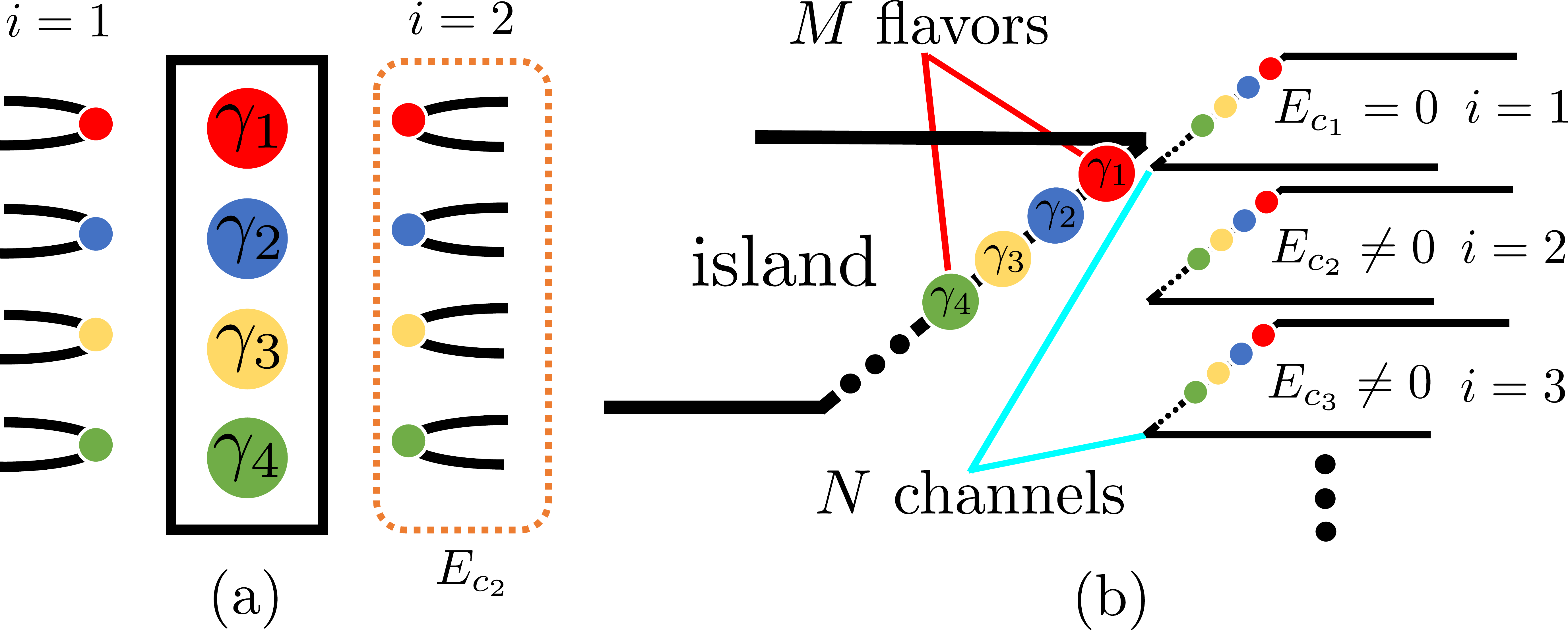}
\caption{(a) $N=2$ $SO(4)$ topological Kondo model. The box at the middle is the Majorana island and the wires (leads) at the left and right hold conduction electrons. Each wire inside a channel is connected to a Majorana zero mode (labeled by $\gamma_{1,2,3,4}$) in the island. (b) $N$-channel $SO(M)$ topological Kondo. The number of Majorana zero modes (flavors) is $M$ and the number of layers (channels) is $N$. The charging energy $E_{c_i}$ ($E_{c_1}$ is zero and all others are nonzero) 
is essential for each channel 
to prevent channel mixing. 
}\label{fig:NtopoK}
\end{figure}

The interaction (\ref{eq:V2C}) arises from a tunneling Hamiltonian $\sum_{i,\alpha}t^i_{\alpha}\gamma_{\alpha}\psi^{\dagger}_{i,\alpha}(x_0)+\mathrm{h.c.}$ between the normal metal leads (fermion operators $\psi_{i,\alpha}$) and the Coulomb blockaded Majorana island (Majorana operators $\gamma_{\alpha}$) with  tunneling amplitudes $t^i_{\alpha}$ (which for simplicity we take to be real) and can be realized in the setup depicted in Fig.~\ref{fig:NtopoK}a. We connect each Majorana of the island to two leads, labeled $i=1,2$, which we call \textit{channels}. The $M$ sub-channels in each channel (which is also the number of Majoranas) are dubbed different \textit{flavors}, labeled by~$\alpha$. In order to prevent tunneling from mixing different channels, we have added a charging energy $E_{c_2}$ for the second channel. Thus the $i=2$ ``lead'' should be considered as a large quantum dot with small level spacing but significant charging energy in full analogy with the proposal of Ref.~\cite{PhysRevLett.90.136602}, used to implement the 2CK effect in quantum dots. 
In the weak-tunneling limit, the effective exchange interaction strength is then $\lambda_{i,\alpha \beta} \propto t^i_{\alpha} t^{i}_{\beta}/U_i$ where $U_i$ is the charging energy. 

We will first focus on 2-channel $SO(4)$ topological Kondo model which can be exactly mapped to 4CK. The reason is that we can consider $SO(M=4)$ as two independent ``spins", i.e. $SO(4)\sim SU(2)\times SU(2)$ which allows us to unitarily transform $\psi_{i,\alpha}$ (with channel index  $i=1,2$ and  flavor index $\alpha=1,2,3,4$) to $\psi_{n,\sigma}\ (n=1,2,3,4;\ \sigma=\uparrow,\downarrow)$ ~\cite{SM}. 
Likewise, with the total charge of the Majorana island fixed, one can form a single $SU(2)$ spin-1/2 out of the four Majorana operators  $\gamma_{\alpha} $. Thus, we have an overscreened Kondo problem which makes the strong coupling fixed point unstable~\cite{nozieres1980kondo} and we expect a stable intermediate coupling fixed point in the isotropic case where we take $\lambda_1 = \lambda_2$ in Eq.~(\ref{eq:V2C}).

Overscreening implies the NFL behavior. 
In order to probe the NFL nature of this low-temperature fixed point, one can measure the fixed point conductance at $T=0$. Due to the charging energy of other channels except the channel $i=1$, we define the conductance matrix~\cite{PhysRevB.96.205403}  $G_{\alpha,\beta}$ in the first channel as the charge current in (flavor) lead $\alpha$ as a linear response to weak voltage $V_\beta\to 0$ applied to lead $\beta$, i.e., $G_{\alpha,\beta} =  \langle I_{\alpha} \rangle / V_\beta$. We expect a nonzero fixed point conductance and the corresponding correction to conductance near $T=0$ will be $T^{(M-2)/(M+2)}$ based on our large-$N$ results and the scaling dimension of the leading irrelevant operator in the CFT~\cite{doi:10.7566/JPSJ.90.024708,affleck1990current,affleck1991kondo,affleck1991critical}, see discussion below Eq.~(\ref{eq:largeNconduc}).
For example, when $N=2$ and $M=4$, we expect 
\begin{equation}
G_{\alpha\neq\beta}(T) = \frac{e^2}{4h}\left[1 +  c_{\alpha\beta} \left(\frac{T}{T_K}\right)^{1/3}\right].\label{eq:2CSO4}
\end{equation}
The  dimensionless coefficients $c_{\alpha\beta}$ (of order one)  and the Kondo temperature $T_K$ are not predicted by the CFT method. 
We used the fact that the 2-channel $SO(4)$ topological Kondo model can be mapped to the 4CK model after fixing the parity~\cite{SM}, see also Eq.~(\ref{eq:Nso4}).  
The nontrivial fractional power of the temperature dependence signifies the NFL behavior at low temperatures. 

Another observable that also shows NFL behavior is the impurity entropy at the fixed point. It is given by $S_{\text{imp}}=\ln g$ where $g$ is usually interpreted as ground state degeneracy. The $N=1$ topological Kondo was shown to have $g=\sqrt{M}$ for odd $M$ and $g=\sqrt{M/2}$ for even $M$ by Altland et al.~\cite{Altland_2014}. Even though the 1-channel impurity entropy shows nontrivial result,  the $N=2$ case is even more complex:
\begin{equation}
g=\begin{cases}
\frac{1}{2}\sqrt{M+2}/\cos\big[\frac{\pi M}{2(M+2)}\big], & M\,\text{is odd,}\\
\frac{1}{2}\sqrt{(M+2)/2}/\cos\big[\frac{\pi M}{2(M+2)}\big], & M\,\text{is even.}
\end{cases} \label{eq:gN2}
\end{equation}
Again, the impurity entropy of 2-channel $SO(4)$ topological Kondo from Eq.~(\ref{eq:gN2}) is $\ln\sqrt{3}$ which is the same as the impurity entropy of 4CK \cite{affleck1995conformal,doi:10.7566/JPSJ.86.084703,doi:10.7566/JPSJ.90.024708} and exhibits the quantum dimension $\sqrt{3}$ of a $Z_3$ parafermion \cite{alicea2015topological}.
Interestingly, 2-channel $SO(8)$ has $g=2+\varphi$, which  indicates an emergent Fibonacci anyon (similar to  3CK). 
Both of the above two observables at this fixed point show fractional values, which are beyond Fermi liquid description and indicate emergent anyonic excitations.
We point out that for a fixed $M$, the impurity entropy and $g$ are larger in the 2-channel case as compared to $N=1$. Indeed,~$g$ approaches monotonically the degeneracy of $M$ free Majoranas in the large-$N$ limit, see Eq.~(\ref{eq:largeNimpurity}) below.

The charging energy $E_{c_2}$ is essential for the multichannel topological Kondo. Otherwise, when $E_{c_2}=0$, the interaction couples only to a single effective channel with a fermion operator $\Tilde{\psi}_{\alpha}(x_0)= \sum_i (t^i_{\alpha}/\abs{\vec{t}_{\alpha}})\psi_{i,\alpha}(x_0)$ and reduces to the conventional $N=1$ topological Kondo Hamiltonian; 
When $E_{c_2}\neq 0$ with fine-tuned $\lambda_1=\lambda_2$, we will have an intermediate coupling fixed point. 
In the case $\lambda_1 \neq \lambda_2$, based on our large-$N$ calculations discussed below, we expect that the weaker coupling renormalizes  to zero and the single-channel limit is recovered. 

\textit{$N\gg 1$: layered construction.}
Next, we generalize the Hamiltonian~(\ref{eq:V2C}) to $N$ channels.
Without exchange isotropy, the interaction becomes
\begin{align}
H_K= \sum_{i=1}^{N}\sum_{\alpha < \beta =2}^M \lambda_{i,\alpha\beta} S^{(\alpha,\beta)} \cdot J_i^{(\alpha,\beta)}(x_0). \label{eq:V}
\end{align}
The coupling constant $\lambda_{i,\alpha \beta}$ is real with symmetry $\lambda_{i,\alpha \beta}=\lambda_{i,\beta\alpha}$ which makes Eq.~(\ref{eq:V}) Hermitian. 
A physical realization for this $N$-channel model can be implemented by using a layered structure depicted in Fig.~\ref{fig:NtopoK}b, where 
 each layer with $M$ flavors encodes a  single $SO(M)$ channel. Here,   channel mixing is prevented by a charging energy $E_{c_i}$ of each channel (except $i=1$)~\cite{PhysRevLett.90.136602}. 
Therefore, $E_{c_i}$ needs to be larger than the temperature or bias voltage (however, large $E_{c_i}$ will decrease the bare Kondo couplings $\lambda_{i,\alpha\beta}$). When $E_{c_i} \neq 0$ with $\lambda$ independent of both channel and flavor indices, we will have an intermediate fixed point at weak coupling, which is found to be stable in terms of anisotropy of flavors; Without fine-tuning of channel couplings, the weaker channel couplings flow to zero, considering large~$N$.

\begin{figure}[tb]
\centering
\includegraphics[width=0.98\columnwidth]{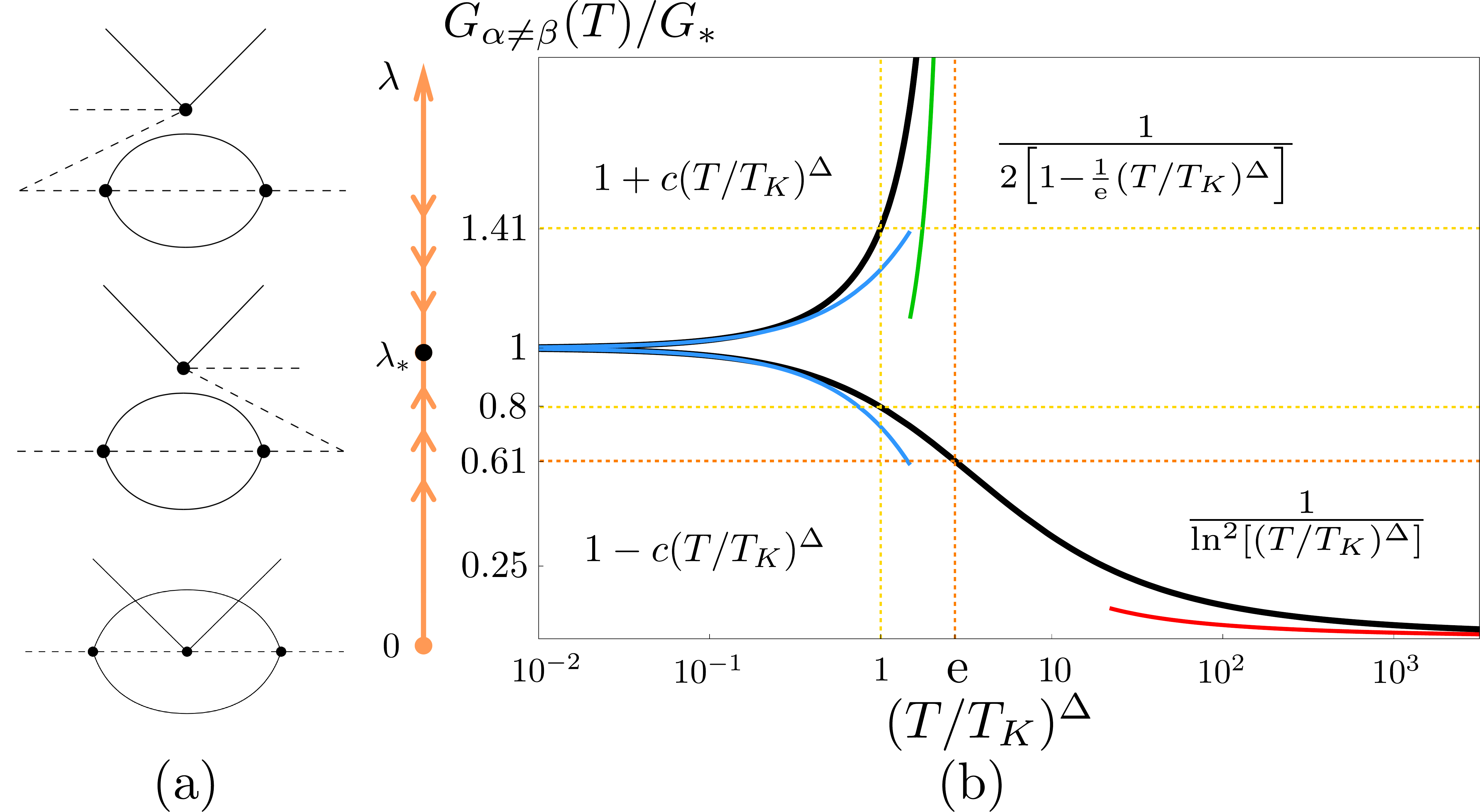}
\caption{(a) The leading Feynman diagrams contributing to the third order RG equation [Eq.~(\ref{eq:beta3})] in the large-$N$ limit. The solid lines denote fermions and dashed lines denote Majorana operators. 
(b)~Left: RG flow of the isotropic Kondo exchange coupling with a stable fixed point $\lambda_*$ (black point). 
~Right: The  conductance vs the temperature for two initial conditions $\lambda_0 \lessgtr \lambda_*$ in the large-$N$ limit. When $\lambda_0<\lambda_*$, the conductance ratio $G(T)/G_*$ is given by the lower black line with a high-temperature approximation $1/\ln^2[(T/T_K)^\Delta]$ (red line) while at low temperature $G(T)/G_* \approx 1 - c (T/T_K)^\Delta $ (lower blue line), with $c = 2/\mathrm{e}^2 \approx 0.27$. 
 Here, $G_*=(e^2/h)(\pi^2/4N^2)$ is the conductance at the fixed point $\lambda_*$, see Eq.~(\ref{eq:largeNconduc}). When $\lambda_0>\lambda_*$, we have $G(T)/G_* \approx 1 + c (T/T_K)^\Delta $ at low temperature (upper blue line) while at high temperature $G$ diverges as $G(T)/G_* \approx 1/[2-2(D/T_K)^\Delta/\mathrm{e}]$ (green line) at  $(T/T_K)^\Delta=\mathrm{e}$, signifying breakdown of the weak coupling perturbation theory.} 
\label{fig:RG}
\end{figure}

By using perturbative renormalization group in the large-$N$ limit~\cite{Anderson_1970,PhysRevLett.70.686,gan1994multichannel,Kogan_2021,kuramoto1998perturbative}, we derive the third order equation for the coupling constant $\lambda_{i,\alpha \beta}$ from Eq. (\ref{eq:V}), 
\begin{eqnarray}
\dv{\lambda_{i,\alpha\beta}}{l}&=&\rho_0(\lambda_i^2)_{\alpha\beta} \label{eq:beta3}\\
&&-\rho_0^2\lambda_{i,\alpha\beta}\sum_j\qty[(\lambda_j^2)_{\alpha\alpha}+(\lambda_j^2)_{\beta\beta}-2(\lambda_{j,\alpha\beta})^2], \nonumber
\end{eqnarray}
where $l=\ln(D_0/D)$ with $D\ (D_0)$ denoting the running (bare)  cutoff energy scale and $\rho_0$ is the density of states per length. 
The three terms at the second line in Eq.~(\ref{eq:beta3}) correspond to the three Feynman diagrams in Fig.~\ref{fig:RG}a. 
 On the isotropic line $\lambda_{i,\alpha\beta}=\lambda$ (for $\alpha \neq \beta$) we have, 
\begin{equation}
    \dv{\lambda}{l}= (M-2)(1-2N\rho_0\lambda)\rho_0 \lambda^2\,. \label{eq:RG_isotropic}
\end{equation}
Thus, the stable intermediate fixed point is $\lambda_{*}=1/(2N\rho_0)$ and moves to weak coupling in the large-$N$ limit. At $\lambda \approx \lambda_{*}$, $ \frac{d \lambda}{dl} \equiv \beta(\lambda)  \approx -(M-2)(\lambda-\lambda_{*}) / (2N)$. The slope of the beta function is $\beta'(\lambda_*) = -(M-2)/(2N)$ which means a large-$N$ scaling dimension  $1+(M-2)/(2N)$ in the irrelevant direction. This agrees with the scaling dimension of the leading irrelevant operator with $N$ channels and $M$ flavors $\Delta_{\text{LIO}} = 1+(M-2)/(2N+M-2)$ obtained from the  CFT~\cite{doi:10.7566/JPSJ.90.024708,affleck1990current,affleck1991kondo,affleck1991critical,PhysRevB.45.7918}~\footnote{The leading irrelevant operator is the first descendent of the adjoint primary operator}. 
The forth and fifth order correction to Eq.~(\ref{eq:RG_isotropic}) are respectively of order $NM^2\rho_0 ^3\lambda^4$ and $N^2M^2\rho_0^4\lambda^5$ and are subleading by a factor $M/N$,
which makes   the large-$N$ perturbation expansion convergent~\cite{PhysRevLett.109.156803,kuramoto1998perturbative}. 

The solution of Eq.~(\ref{eq:RG_isotropic}) is 
\begin{equation}
    \frac{\lambda(D)}{\lambda_*}=f^{-1}\qty[\frac{(D/T_K)^\Delta}{\me^2}], \ f(x)=|1/x-1|\me^{1/x-1},\label{eq:solisoRG}
\end{equation}
where we introduced the Kondo temperature $T_K=D_0 [\me^2f(\lambda_0/\lambda_*)]^{-1/\Delta}$ and $\Delta=(M-2)/(2N)$. 
We will have two solutions for Eq.~(\ref{eq:solisoRG}) depending on whether the initial (bare) coupling constant $\lambda_0=\lambda(D_0)$ is larger or smaller than the fixed point coupling $\lambda_*$ (see Fig. \ref{fig:RG}b). 
Since the low-energy fixed point is at weak coupling in the large-$N$ limit, Eq.~(\ref{eq:solisoRG}) gives the full cross over for the running coupling $\lambda(D)$, extending beyond the CFT prediction.

\textit{Conductance at large-$N$.}
In both $N=1$, $SO(M)$ topological Kondo \cite{PhysRevLett.109.156803,PhysRevResearch.2.043228,PhysRevLett.110.216803,PhysRevB.94.235102,PhysRevB.96.205403} and $N=2$, $SO(4)$ topological Kondo [Eq.~(\ref{eq:2CSO4})], the fixed point ($T=0$) conductance is given by a universal fractional  multiple of $G_0=e^2/h$.
Here, we first evaluate the conductance perturbatively in $\lambda_0$ by using Kubo formula~\cite{SM} and find $G_{\alpha\beta}=G_0(\pi\lambda_0  \rho_0)^2 (1 -M \delta_{\alpha\beta})$ to lowest order. By evaluating the next-order correction to the conductance at finite frequency $\omega$, we find a logarithmic divergence $\sim \lambda_0^3 \ln (D/D_0)$ where $D = \max \{T,\omega \}$. The divergence results from renormalization of the coupling $\lambda$, described by the RG equation~\eqref{eq:RG_isotropic}. 
This indicates that for $T\gg \omega$,  $G_{\alpha\neq\beta}(T)=G_0[\pi\lambda(T) \rho_0]^2$, plotted 
as a function of temperature in  Fig.~\ref{fig:RG}b. Remarkably, in the large-$N$ limit $\lambda(T)$ remains small and the full cross over function, Eq.~(\ref{eq:solisoRG}), can be found exactly as long as the bare coupling $\lambda_0$ is small.
At low temperatures, $T \ll T_K$, the conductance approaches its zero-temperature value with a power-law characteristic of a NFL, 
\begin{equation}
G_{\alpha\neq\beta} (T)/G_0 =\frac{\pi^2}{4N^2} \left[1 + c_{\alpha \beta}  \left(\frac{T}{T_K}\right)^{(M-2)/2N}\right],\label{eq:largeNconduc}
\end{equation} 
where the dimensionless constant can be explicitly obtained in the isotropic case, $c_{\alpha \beta} = \pm \delta_{\alpha, \beta} c$ with $c =  \frac{2}{\me^2} \approx 0.27$, based on the cross over function, Eq.~(\ref{eq:solisoRG}). 
(The sign $\pm$ is determined by the initial condition $\lambda_0 \gtrless \lambda_*$.)  
The temperature-dependent correction $\sim T^{\Delta_{\text{LIO}} - 1}$, also obtained from Eq.~(\ref{eq:solisoRG}), matches with first-order correction from the leading irrelevant operator, see below Eq.~(\ref{eq:RG_isotropic}), somewhat similar to the case of resistivity in $SU(2)$ MCK~\cite{PhysRevB.48.7297}. 
This is notably different from the single-channel topological Kondo effect where the first order correction vanishes~\cite{PhysRevLett.109.156803} and temperature correction is $ \sim T^{2(\Delta_{\text{LIO}} - 1)}$.
 
The fixed point ($T=0$) conductance above (i.e., the first term) can be verified for $M=4$, in which case we can map the $N$-channel $SO(4)$ topological Kondo to $2N$CK by a unitary transformation~\cite{SM}. 
From the mapping, we find the conductance of the $N$-channel $SO(4)$ model~\cite{SM}: 
\begin{equation}
G_{\alpha\neq\beta}(M=4,T=0)/G_0 = \sin^2[\pi/(2N+2)] \ ,\label{eq:Nso4}
\end{equation}
which bears resemblance to the $2N$CK fixed point  conductance~\cite{PhysRevB.65.195101,PhysRevB.57.R5579}. 
The $N=2$ case gives the fixed point conductance (first term) in Eq. (\ref{eq:2CSO4}). The $N=1$ result agrees with Ref.~\cite{PhysRevLett.110.216803,PhysRevB.94.235102,PhysRevB.96.205403} while the large-$N$ limit agrees with our previous result,  Eq.~(\ref{eq:largeNconduc}).

\textit{Impurity entropy at large-$N$.} 
The NFL nature of the low-temperature fixed point becomes apparent in the impurity entropy $ S_{\text{imp}} = \ln g $ where $g$ can take a non-integer value. As mentioned in the introduction and displayed by Eq.~(\ref{eq:gN2}), the 1 and 2-channel topological Kondo models generically show a non-integer $g$. Also, $g$ for the $N$-channel case can be calculated by using modular S-matrix~\cite{doi:10.7566/JPSJ.90.024708, PhysRevB.58.3794}. The modular S-matrix of $SO(M)$ is given in Ref.~\cite{hung2018linking} and the general information of it can also be found in Ref.~\cite{francesco2012conformal}. We then find Eq.~(\ref{eq:gN2}) in the case $N=2$. Above we saw that in the large-$N$ limit the fixed point coupling moves to weak coupling. In this case the impurity is weakly screened and we find in the large-$N$ limit,
\begin{align}
g=\begin{cases}
2^{(M-1)/2}\big[1-\frac{(M-2)(M-1)M\pi^{2}}{192N^{2}}\big], & M\,\text{is odd}\\
2^{(M-2)/2}\big[1-\frac{(M-2)(M-1)M\pi^{2}}{192N^{2}}\big], & M\,\text{is even.}
\end{cases}\label{eq:largeNimpurity}
\end{align}
This result indeed reflects the fact that the impurity is almost free at large-$N$ [same conclusion can be made from the conductance~(\ref{eq:largeNconduc})]. The impurity entropy is that of $M$ free Majoranas (with a fixed total parity) with a correction of order $1/N^2$ from the screening by the itinerant electrons.
The values in Eq. (\ref{eq:gN2}) are the ground state degeneracy at the fixed point $\lambda_*$ without taking large-$N$ limit. They are smaller than the above first term, giving the ground state degeneracy at $\lambda=0$. This agrees with the $g$-theorem in CFT,  stating that the ground state degeneracy becomes smaller along the RG flow \cite{PhysRevLett.67.161,PhysRevB.48.7297,doi:10.7566/JPSJ.86.084703,PhysRevLett.93.030402} from $\lambda = 0$ to $\lambda_*$.

\begin{figure}[tb]
\centering
\includegraphics[width=0.98\columnwidth]{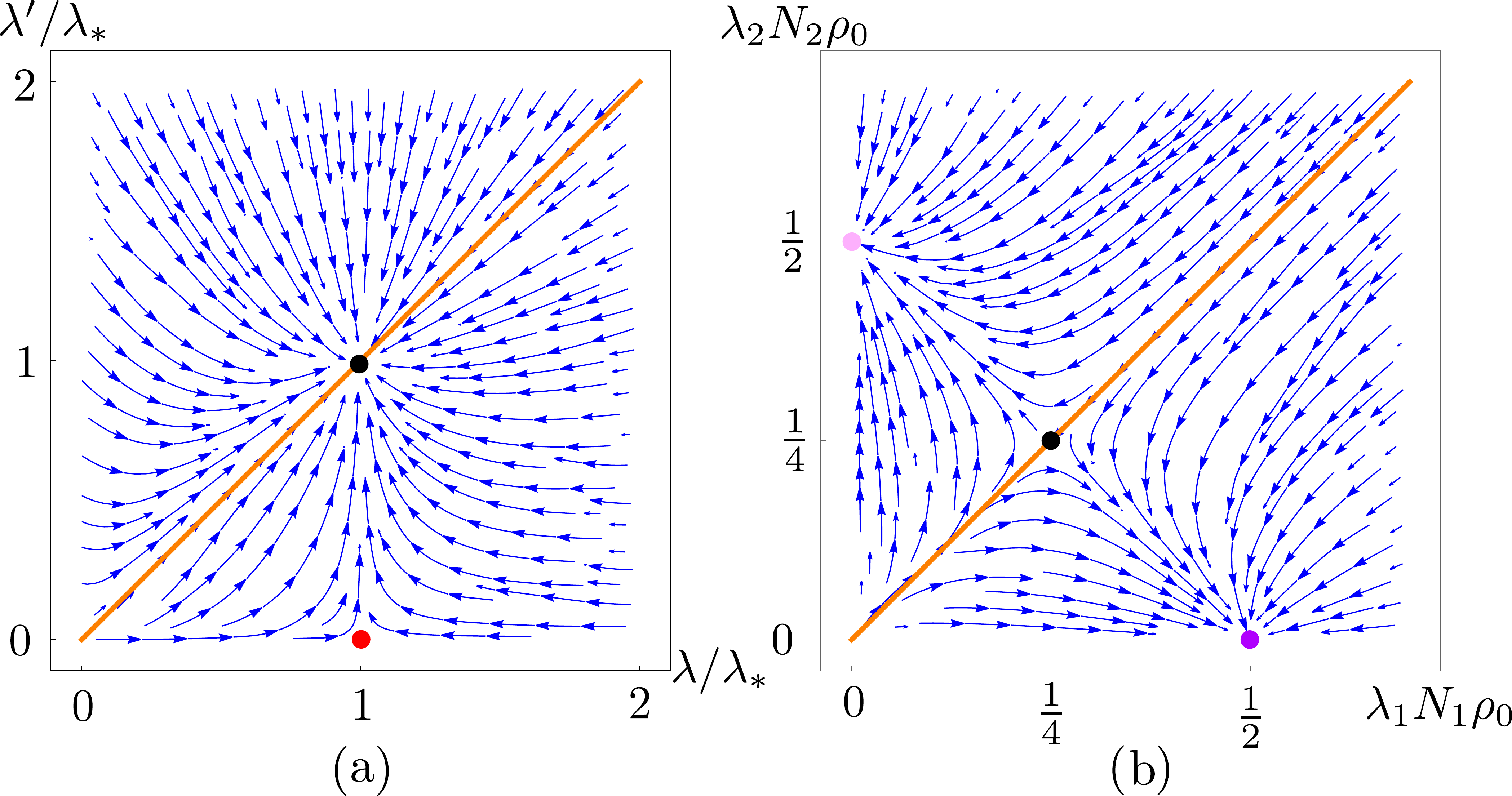}
\caption{RG flow of (a) flavor and (b) channel anisotropies. The  isotropic lines are denoted in orange with the black point showing the isotropic intermediate fixed point. (a)~Flavor anisotropy. The isotropic fixed point (black point) is at $\lambda=\lambda'=1/(2N\rho_0)$ and is stable. The red fixed point is the isotropic fixed point of the $SO(M-1)$ model and  is unstable and anisotropic in terms of $M$ flavors. (b)~Channel anisotropy with the  case $N_1=N_2$ shown. The isotropic fixed point (black point) is at $\lambda_1=\lambda_2 =1/2[(N_1+N_2)\rho_0]$ which is unstable.
Depending on which of $\lambda_1$ and $\lambda_2$ is larger,  the stable fixed point is the purple [$\lambda_1=1/(2N_1\rho_0),\lambda_2=0$] or the pink [$\lambda_1=0,\lambda_2=1/(2N_2\rho_0)$] one. } 
\label{fig:anisotropy}
\end{figure}

\textit{Flavor anisotropy. }
Since the NFL behavior in the conventional single-channel topological Kondo model is stable to flavor anisotropy \cite{PhysRevLett.109.156803}, it is natural to expect the same to be true for its multichannel generalization. 
Indeed, by considering the physically-motivated~\cite{SM} flavor-anisotropic coupling  $\lambda_{\alpha\beta}=[\lambda+(\lambda'-\lambda)(\delta_{1\alpha}+\delta_{1\beta})]$ in Eq.~\eqref{eq:beta3}, 
we find  two nontrivial fixed points with the majority coupling $\lambda =1/(2N\rho_0)$ in both, while the minority coupling is either $\lambda' =1/(2N \rho_0)$ or $\lambda' =0$. 
The first one is isotropic and stable, while the    second fixed point is unstable, see Fig. \ref{fig:anisotropy}a. 
Thus, flavor anisotropy remains irrelevant  in the multichannel generalization of the topological Kondo model.

\textit{Channel anisotropy.}
Since the NFL fixed point of the conventional $SU(2)$ MCK is unstable to channel anisotropy~ \cite{nozieres1980kondo,PhysRevB.45.7918,PhysRevLett.74.4507}, it is crucial to investigate channel anisotropy in the multichannel topological Kondo model. 
We consider a flavor isotropic but channel anisotropic version of   Eq.~(\ref{eq:V}) with $\lambda_{i,\alpha\beta}=\lambda_1$ for $i=1,\dots,N_1$ and $\lambda_{N_1 + i,\alpha\beta}=\lambda_2$ for $i=1,\dots,N_2$. If we consider large-$N_1$ and large-$N_2$ limit, we will have ($j=1,2$)
\begin{equation}
\dv{\lambda_j}{l}=(M-2)\lambda_j\rho_0(\lambda_j-2N_1\lambda_1^2 \rho_0-2N_2\lambda_2^2 \rho_0)\,.\label{eq:channelaniso}   
\end{equation} 
The RG flow is shown in Fig.~\ref{fig:anisotropy}b and has 2 stable anisotropic fixed points. When $\lambda_1 \neq \lambda_2$, the smaller coupling constant flows to zero and the larger one, $\lambda_i$,  flows to
$1/(2N_i \rho_0 )$. The isotropic fixed point  $\lambda_1=\lambda_2 =1/[2(N_1+N_2)\rho_0]$ is  unstable.

\textit{Conclusions.} 
We generalized the topological Kondo interaction into its $N\geq2$-channel version by adding new sets of floating leads connected to the Majorana island (see Fig.~\ref{fig:NtopoK}). Consequently, we analyzed the two-channel case for its 
impurity entropy [Eq.~(\ref{eq:gN2})] and  conductance [Eq.~(\ref{eq:2CSO4})]. The former indicates an emergent Fibonacci anyon, beyond the single-channel model. 
Another departure from the single-channel model is the NFL correction to conductance, which is of first order in the irrelevant operator, see below Eq.~(\ref{eq:largeNconduc}). 
The introduced multichannel generalization allowed us to develop a convergent large-$N$ perturbation theory.  
Under the large-$N$ limit, we found that the multichannel topological Kondo has a stable fixed point at weak coupling. By using perturbative RG, we were able to solve for the running coupling constant, giving us the full cross over from the free fixed point to the intermediate one, see Eq.~(\ref{eq:solisoRG}) and Fig.~\ref{fig:RG}. 
We also considered the flavor and channel anisotropies in the large-$N$ limit [see Fig. \ref{fig:anisotropy}, Ref. \cite{SM} and Eqs.~(\ref{eq:channelaniso})], finding that the flavor anisotropy is irrelevant, while the channel anisotropy is relevant. Our work suggests further study into the exotic physics found in multichannel Kondo models that are beyond conventional, $SU(2)$-symmetric, spin systems.  
 
\textit{Acknowledgments.}
We thank Sergei Khlebnikov and Elio K\"onig for valuable discussions. This work was initiated at the Aspen Center for Physics, which is supported by National Science Foundation grant PHY-1607611. Y.O. acknowledges support by the European Union’s Horizon 2020 research and innovation programme (Grant Agreement LEGOTOP No. 788715), the DFG (CRC/Transregio 183, EI 519/7-1), ISF Quantum Science and Technology (2074/19), the BSF and NSF (2018643). This material is based upon work supported by the U.S. Department of Energy, Office of Science, National Quantum Information Science Research Centers, Quantum Science Center. 

\bibliography{apssamp}

\clearpage
\newpage

\setcounter{equation}{0}
\setcounter{figure}{0}
\setcounter{section}{0}
\setcounter{table}{0}
\setcounter{page}{1}
\makeatletter
\renewcommand{\theequation}{S\arabic{equation}}
\renewcommand{\thepage}{S-\arabic{page}}
\renewcommand{\thesection}{S\arabic{section}}
\renewcommand{\thefigure}{S\arabic{figure}}
\renewcommand{\thetable}{S-\Roman{table}}

\begin{widetext}

\section*{Supplementary Material to ``Multichannel topological Kondo effect''}

In this Supplementary Material, we present additional detail on our calculations of large-$N$ fixed point conductance, the equivalence between the conventional $SU(2)$ $2N$CK and the $N$-channel $SO(4)$ topological Kondo and use this to evaluate the fixed point conductance for the $SO(4)$ topological Kondo model. We also show the impurity entropy for general $M, N$, deriving its large-$N$ limit, explain the details of the flavor anisotropy we considered  and discuss channel anisotropy in the case of $\mathcal{O}(1)$ anisotropic channels.


\section{Conductance in the weak-tunneling limit} 
In this section, we evaluate the conductance in the limit of large number of channels $N$. We will use the Kubo formula: 
\begin{equation}
G_{\alpha\beta}^{ij}=  \Re\lim_{\omega\to0}\frac{-1}{\hbar\omega}\int_{0}^{\infty}\mathrm{\mathrm{d}}t \, \mathrm{e}^{\mathrm{i}\omega_{+}t}\langle[I_{\alpha}^{i}(t),I_{\beta}^{j}(0)]\rangle , \label{eq:Kubo}
\end{equation}
where $\omega_{+}=\omega+\mi\eta$ with $\eta$ a positive infinitesimal. 
For it, we will need to first evaluate the current-current correlation function, which we now derive.  
In the weak-tunneling limit, the charge current operator for  channel $j$ and flavor $\rho$ is most conveniently defined as 
\begin{equation}
I_{j,\rho}=-e\dv{N_{j,\rho}}{t}=-\mi\frac{e}{\hbar}[H,N_{j,\rho}],\ \text{where}\ N_{j,\rho}=\int_{-\infty}^{\infty}\dd{x}\psi_{j,\rho}^{\dagger}(x)\psi_{j,\rho}(x),
\end{equation}
and the integration is over the lead. In the main text, we take $j=1$, i.e., we consider the channel that is not floating (has no charging energy term).
Generally, the full Hamiltonian $H$ can be written as 
\begin{equation}
H=H_{0}+H_K=\sum_{i=1}^{N}\sum_{\alpha=1}^{M}\sum_{p}\epsilon_{p}c_{p,i\alpha}^{\dagger}c_{p,i\alpha}-\text{\ensuremath{\frac{1}{2}}}\sum_{i,k=1}^{N}\sum_{\alpha\neq\beta=1}^{M}\lambda_{\alpha\beta}^{ik}\gamma_{\alpha}\gamma_{\beta}\psi_{i,\alpha}^{\dagger}(0)\psi_{k,\beta}(0), 
\end{equation}
and we allow for generic anisotropies in the interaction term. Hermiticity  requires that $\lambda_{\alpha\beta}^{ik}=\lambda_{\beta\alpha}^{ki}$.
Since $H_{0}$ conserves the particle number in a lead, the current is
given by the commutator of $\hat{N}_{j,\rho}$ and $V$,  
\begin{align}
I_{j,\rho}=-\mi\frac{e}{\hbar}[H_K,N_{j,\rho}]=\mi\frac{e}{2\hbar}\sum_{i=1}^{N}\sum_{\alpha=1}^{M}\lambda_{\alpha\rho}^{ij}\gamma_{\alpha}\gamma_{\rho}\qty[\psi_{i,\alpha}^{\dagger}(0)\psi_{j,\rho}(0)+\psi_{j,\rho}^{\dagger}(0)\psi_{i,\alpha}(0)]
\end{align}
where we used the Hermiticity requirement to combine two sums. Then,
we can find the following average 
\begin{align}
\langle I_{\alpha}^{i}(t)I_{\beta}^{j}(0)\rangle=-\frac{e^{2}}{2\hbar^{2}}\big[(\lambda_{\alpha\beta}^{ij})^{2}-\sum_{i',\alpha'}\lambda_{\alpha'\alpha}^{i'i}\lambda_{\alpha'\beta}^{i'j}\delta_{ij}\delta_{\alpha\beta}\big]G_{<}(-t)G_{>}(t).
\end{align}
Here, we define the Green's functions,  
\begin{equation}
\langle\psi_{i',\alpha'}^{\dagger}(t)\psi_{j,\beta}(0)\rangle=-\mathrm{i}G_{<}(-t)\delta_{i'j}\delta_{\alpha'\beta}\ \text{and}\ \langle\psi_{i,\alpha}(t)\psi_{j',\beta'}^{\dagger}(0)\rangle=\mathrm{i}G_{>}(t)\delta_{ij'}\delta_{\alpha\beta'}.
\end{equation}
By Fourier transforming into momentum space, we get for free fermions 
\begin{align}
G_{<}(-t)= & \frac{\mathrm{i}}{L}\sum_{p}\mathrm{e}^{\mathrm{i}\epsilon_{p}t/\hbar}f(\epsilon_{p})=\mathrm{i}\mathrm{\rho_{0}}\int_{-\infty}^{0}\mathrm{d}\epsilon\,\mathrm{e}^{\mathrm{i}\epsilon(t-\mathrm{i}\eta)/\hbar}=\frac{\hbar\rho_{0}}{t-\mathrm{i}\eta},\\
G_{>}(t)= & -\frac{\mathrm{i}}{L}\sum_{p'}\mathrm{e}^{-\mathrm{i}\epsilon_{p'}t/\hbar}[1-f(\epsilon_{p'})]=-\mathrm{i}\mathrm{\rho_{0}}\int_{0}^{\infty}\mathrm{d}\epsilon\,\mathrm{e}^{-\mathrm{i}\epsilon(t-\mathrm{i}\eta)/\hbar}=-\frac{\hbar\rho_{0}}{t-\mathrm{i}\eta}\,,
\end{align}
where $f(\epsilon)=(\mathrm{e}^{\beta\epsilon}+1)^{-1}$ is the Fermi function,  replaced by a step function at $T=0$.  
Also, we approximated the density of states as $\nu(E) \approx \nu(E_F)$. We defined $\rho_0=\nu(E_F)/L$ as the density of states per length at the Fermi energy. 
We have regularized the integration over energy by considering using the short-time (UV) cutoff $\eta=\hbar/D_0$ 
($D_0$ is the bare cutoff mentioned in the main text). 
We then get 
\begin{align}
G_{<}(-t)G_{>}(t) & =-\hbar^{2}\rho_0^2\frac{1}{(t-\mathrm{i}\eta)^{2}};\ \
\langle I_{\alpha}^{i}(t)I_{\beta}^{j}(0)\rangle  =2e^{2}\rho_0^2\big[(\lambda_{\alpha\beta}^{ij})^{2}-\sum_{i',\alpha'}\lambda_{\alpha'\alpha}^{i'i}\lambda_{\alpha'\beta}^{i'j}\delta_{ij}\delta_{\alpha\beta}\big]\frac{1}{(t-\mathrm{i}\eta)^{2}}.
\end{align}
By using 
\begin{equation}
\mathrm{Im}\lim_{\eta\to0^{+}}-\frac{1}{(t-\mathrm{i}\eta)^{2}}=\mathrm{Im}\lim_{\eta\to0^{+}}\partial_{t}\frac{1}{t-\mathrm{i}\eta}=\partial_{t}\lim_{\eta\to0^{+}}\frac{\eta}{t^{2}+\eta^{2}}=\pi\partial_{t}\delta(t),
\end{equation}
and find 
\begin{equation}
G_{<}(-t)G_{>}(t)-G_{<}(t)G_{>}(-t)=2\mathrm{i}\,\mathrm{Im}[G_{<}(-t)G_{>}(t)]=2\pi\mathrm{i}\hbar^{2}\rho_0^2\partial_{t}\delta(t).
\end{equation}
Thus, we have 
\begin{align}
\langle[I_{\alpha}^{i}(t),I_{\beta}^{j}(0)]\rangle= & -\frac{e^{2}}{4\hbar^{2}}\sum_{i',j'}\sum_{\alpha',\beta'}\lambda_{\alpha'\alpha}^{i'i}\lambda_{\beta'\beta}^{j'j}\times2[G_{<}(-t)G_{>}(t)-G_{<}(t)G_{>}(-t)](\delta_{i'j}\delta_{ij'}\delta_{\alpha'\beta}\delta_{\alpha\beta'}-\delta_{ij}\delta_{i'j'}\delta_{\alpha\beta}\delta_{\alpha'\beta'})\\
= & -\mathrm{i}\pi e^{2}\rho_0^2\partial_{t}\delta(t)\big[(\lambda_{\alpha\beta}^{ij})^{2}-\sum_{i',\alpha'}\lambda_{\alpha'\alpha}^{i'i}\lambda_{\alpha'\beta}^{i'j}\delta_{ij}\delta_{\alpha\beta}\big]
\end{align}
where we used   $\lambda_{\alpha\beta}^{ij}=\lambda_{\beta\alpha}^{ji}$
and 
\begin{align}
& \langle[\gamma_{\alpha'}\gamma_{\alpha}\big(\psi_{i',\alpha'}^{\dagger}(t)\psi_{i,\alpha}(t)+\psi_{i,\alpha}^{\dagger}(t)\psi_{i',\alpha'}(t)\big),\gamma_{\beta'}\gamma_{\beta}\big(\psi_{j',\beta'}^{\dagger}(0)\psi_{j,\beta}(0)+\psi_{j,\beta}^{\dagger}(0)\psi_{j',\beta'}(0)\big)]\rangle\\
= & 2[G_{<}(-t)G_{>}(t)-G_{<}(t)G_{>}(-t)](\delta_{i'j}\delta_{ij'}\delta_{\alpha'\beta}\delta_{\alpha\beta'}-\delta_{ij}\delta_{i'j'}\delta_{\alpha\beta}\delta_{\alpha'\beta'}).
\end{align}

We obtain the conductance from the Kubo formula Eq.~(\ref{eq:Kubo}), which gives 
\begin{equation}
G_{\alpha\beta}^{ij}=  -\pi\frac{e^{2}}{\hbar}\rho_0^2\big[(\lambda_{\alpha\beta}^{ij})^{2}-\sum_{i',\alpha'}\lambda_{\alpha'\alpha}^{i'i}\lambda_{\alpha'\beta}^{i'j}\delta_{ij}\delta_{\alpha\beta}\big]\Im\lim_{\omega\to0}\frac{1}{\omega}\int_{0}^{\infty}\mathrm{\mathrm{d}}t \, \mathrm{e}^{\mathrm{i}\omega_{+}t}\partial_{t}\delta(t)
\end{equation} 
The imaginary part of the integral above is 
\begin{align}
\Im\lim_{\omega\to0}\frac{1}{\omega}\int_{0}^{\infty}\mathrm{\mathrm{d}}t\, \mathrm{e}^{\mathrm{i}\omega_{+}t}\partial_{t}\delta(t) & =\Im\lim_{\omega\to0}\frac{1}{\omega}\big[\mathrm{e}^{\mathrm{i}\omega_{+}t}\delta(t)\big]_{0}^{\infty}-\Im\lim_{\omega\to0}\frac{\mathrm{i}\omega_{+}}{\omega}\int_{0}^{\infty}\mathrm{\mathrm{d}}t\, \mathrm{e}^{\mathrm{i}\omega_{+}t}\delta(t)=-\frac{\mathrm{1}}{2} \,.
\end{align} 
Thus, by defining $G_{0}=e^{2}/h$, the real part of the conductance is 
\begin{equation}
G_{\alpha\beta}^{ij}/G_{0}=(\pi\rho_{0})^{2}\big[(\lambda_{\alpha\beta}^{ij})^{2}-\sum_{i',\alpha'}\lambda_{\alpha'\alpha}^{i'i}\lambda_{\alpha'\beta}^{i'j}\delta_{ij}\delta_{\alpha\beta}\big].
\end{equation}
For the isotropic case $\lambda_{\alpha\beta}^{ij}=\lambda\delta_{ij}(1-\delta_{\alpha\beta})$, we find 
\begin{equation}
G_{\alpha\beta}^{ii}/G_{0}=  (\pi\lambda\rho_{0})^{2} (1-M\delta_{\alpha\beta}).\label{eq:isoconductance}
\end{equation}
In the large-$N$ limit, we can access the intermediate fixed point
within perturbation theory. In this case, with fixed point coupling
$\lambda_{*}=1/(2N\rho_{0})$, we find a zero-temperature  conductance $G_{\alpha\beta}^{ii}/G_{0}=  (\pi^{2}/4N^{2})(1-M\delta_{\alpha\beta})$, which gives the first term in Eq.~(\ref{eq:largeNconduc}) of the main text. 
Next, we argue that in the perturbative limit we can use the renormalized coupling $\lambda(D)$ and write  $G_{\alpha\beta}^{ij}(D) \propto \lambda(D)^2$ to obtain the conductance at cutoff scale $D$ (which can be replaced by temperature or frequency), see below Eq.~\eqref{seq:G3rdOrder}.

\subsection{The next-order correction to conductance in the weak coupling}

To go beyond the above discussed lowest order calculation, it is convenient to use the following expression for   
the weak-tunneling correction to the average of operator $O(t)$:
\begin{equation}
\delta\left\langle O(t)\right\rangle =-\frac{\mi}{\hbar}\int_{-\infty}^{t}dt'\left\langle [O(t),H_K(t')]\right\rangle \,,
\end{equation}
where  $H_K$ is the tunneling perturbation. 
Taking $O(t)=[I_{\alpha}^{i}(t),I_{\beta}^{j}(0)]$, we obtain the
next-order correction to conductance [see Eq.~(\ref{eq:Kubo})], 
\begin{equation}
\delta G_{\alpha\beta}^{ij} (\omega) =\Re  \frac{\mi}{\hbar^{2}\omega}\int_{0}^{\infty}\mathrm{\mathrm{d}}t\mathrm{e}^{\mathrm{i}\omega_{+}t}\int_{-\infty}^{t}dt'\left\langle [[I_{\alpha}^{i}(t),I_{\beta}^{j}(0)],H_K(t')]\right\rangle \,.
\end{equation}
(Since we are here considering the $T=0$ limit, we will see later that in the limit $\omega \to 0$ the correction $\delta G_{\alpha\beta}^{ij} (\omega)$  diverges.)  
We note that $[I_{\alpha}^{i}(t),I_{\beta}^{j}(0)]^{\dagger}=-[I_{\alpha}^{i}(t),I_{\beta}^{j}(0)]$.
Thus, 
\begin{flalign}
\left\langle [[I_{\alpha}^{i}(t),I_{\beta}^{j}(0)],H_K(t')]\right\rangle   =\left\langle [I_{\alpha}^{i}(t),I_{\beta}^{j}(0)]H_K(t')\right\rangle -\left\langle ([I_{\alpha}^{i}(t),I_{\beta}^{j}(0)]^{\dagger}H_K(t'))^{\dagger}\right\rangle =\left\langle [I_{\alpha}^{i}(t),I_{\beta}^{j}(0)]H_K(t')\right\rangle +c.c.
\end{flalign}
and 
\begin{equation}
\delta G_{\alpha\beta}^{ii}=\Re\lim_{\omega\to0}\frac{\mi}{\hbar^{2}\omega}\int_{0}^{\infty}\mathrm{\mathrm{d}}t\mathrm{e}^{\mathrm{i}\omega_{+}t}\int_{-\infty}^{t}dt'\left(\left\langle [I_{\alpha}^{i}(t),I_{\beta}^{i}(0)]H_K(t')\right\rangle +c.c.\right)\label{eq:deltaG}
\end{equation}
Now we can insert $I_{j,\rho}$ and $H_K$ and evaluate the average.
Let's take for simplicity $\lambda_{\alpha\rho}^{ij}=\delta_{ij}\lambda_{\alpha\rho}$,
i.e., channel-isotropic exchange. Let us introduce $W_{\alpha\beta}^{i}=W_{\beta\alpha}^{i\dagger}=\gamma_{\alpha}\gamma_{\beta}\psi_{i,\alpha}^{\dagger}(0)\psi_{i,\beta}(0)$. Then 
\begin{equation}
I_{i,\alpha}  =\mi\frac{e}{2\hbar}\sum_{\alpha'=1}^{M}\lambda_{\alpha'\alpha}W_{\alpha'\alpha}^{i}+h.c. =\mi\frac{e}{2\hbar}\sum_{\alpha'=1}^{M}\lambda_{\alpha'\alpha}(W_{\alpha'\alpha}^{i}-W_{\alpha\alpha'}^{i}),\ \
H_K  =-\text{\ensuremath{\frac{1}{2}}}\sum_{j=1}^{N}\sum_{\gamma\neq\delta=1}^{M}\lambda_{\gamma\delta}W_{\gamma\delta}^{j}
\end{equation}
and 
\begin{align}
\left\langle [I_{\alpha}^{i}(t),I_{\beta}^{i}(0)]H_K(t')\right\rangle +c.c. =& 2\qty(\frac{e}{2\hbar})^{2}\text{\ensuremath{\frac{1}{2}}}\sum_{j=1}^{N}\sum_{\gamma\neq\delta=1}^{M}\sum_{\alpha'\beta'=1}^{M}\lambda_{\beta'\beta}\lambda_{\alpha'\alpha}\lambda_{\gamma\delta} \notag \\ 
&\Re\left\langle [W_{\alpha'\alpha}^{i}(t)-W_{\alpha\alpha'}^{i}(t),W_{\beta'\beta}^{i}(0)-W_{\beta\beta'}^{i}(0)]W_{\gamma\delta}^{j}(t')\right\rangle.
\end{align}
Thus,
\begin{align}
 & \left\langle [W_{\alpha'\alpha}^{i}(t)-W_{\alpha\alpha'}^{i}(t),W_{\beta'\beta}^{i}(0)-W_{\beta\beta'}^{i}(0)]W_{\gamma\delta}^{j}(t')\right\rangle =-\mi F_{\alpha\alpha'\beta\beta'\gamma\delta}\delta_{ij}G_{>}(t-t')G_{>}(-t')[G_{>}(t)-G_{>}(-t)]\,,\\
 &
F_{\alpha\alpha'\beta\beta'\gamma\delta}=\delta_{\alpha'\delta}\delta_{\alpha\beta'}\delta_{\beta\gamma}+\delta_{\alpha\gamma}\delta_{\delta\beta'}\delta_{\alpha'\beta}-\delta_{\alpha\delta}\delta_{\alpha'\beta'}\delta_{\beta\gamma}-\delta_{\alpha'\gamma}\delta_{\delta\beta'}\delta_{\alpha\beta}-(\beta\leftrightarrow\beta')\,.
\end{align}
We need the real part of the above average:
\begin{equation}
\Re\left\langle [W_{\alpha'\alpha}^{i}(t)-W_{\alpha\alpha'}^{i}(t),W_{\beta'\beta}^{i}(0)-W_{\beta\beta'}^{i}(0)]W_{\gamma\delta}^{j}(t')\right\rangle =\hbar^{3}\rho_{0}^{3}F_{\alpha\alpha'\beta\beta'\gamma\delta}\delta_{ij}\frac{(2t'-t)2t}{(t-t')^{2}+\eta^2}\frac{\eta}{t^{\prime2}+\eta^2}\frac{1}{t^{2}+\eta^2}.
\end{equation}
The Fourier transform of this function of $t$ is 
\begin{flalign}
I(\omega) & =\int_{0}^{\infty}\mathrm{d}t\mathrm{e}^{\mathrm{i}\omega_{+}t}\int_{-\infty}^{t}dt'\frac{(2t'-t)2t}{(t-t')^{2}+\eta^2}\frac{\eta}{t^{\prime2}+\eta^2}\frac{1}{t^{2}+\eta^2} \approx-\int_{0}^{\infty}\mathrm{\mathrm{d}}t\mathrm{e}^{\mathrm{i}\omega_{+}t}\frac{2t^{2}}{t^{2}+\eta^2}\pi\frac{1}{t^{2}+\eta^2}\Theta(t)\\
 & =-2\pi\int_{0}^{\infty}\mathrm{\mathrm{d}}t\mathrm{e}^{\mathrm{i}\omega_{+}t}\left(1-\frac{\eta^2}{t^{2}+\eta^2}\right)\frac{1}{t^{2}+\eta^2}.
\end{flalign}
For the  conductance, we need the imaginary part of $I$ in the limit $\omega\ll D_0 = \hbar/\eta$, 
\begin{equation}
\Im I(\omega)  =-2\pi\int_{0}^{\infty}\mathrm{\mathrm{d}}t\left(1-\frac{\eta^2}{t^{2}+\eta^2}\right)\frac{\sin\omega t}{t^{2}+\eta^2} \approx-2\pi\int_{0}^{\infty}\mathrm{\mathrm{d}}t\frac{\sin\omega t}{t^{2}+\eta^2} \approx 2\pi \omega\ln(\hbar\omega / D_0),
\end{equation} 
where the subleading terms are of order $\omega \eta$~\cite{Gradshteyn}. 
Then at small $\omega$, we have $\Im I(\omega)/\omega=2\pi\ln(\hbar\omega/D_0)$. After combining the above results with the conductance correction formula Eq.~(\ref{eq:deltaG}), we immediately get
\begin{align}
\delta G_{\alpha\beta}^{ii} (\omega) = & \Re \frac{\mi}{\hbar^{2}\omega}\int_{0}^{\infty}\mathrm{\mathrm{d}}t\mathrm{e}^{\mathrm{i}\omega_{+}t}\int_{-\infty}^{t}dt'\left(\left\langle [I_{\alpha}^{i}(t),I_{\beta}^{i}(0)]H_K(t')\right\rangle +c.c.\right)\\
= &- 2(\frac{e}{2\hbar})^{2}\text{\ensuremath{\frac{1}{2}}}\frac{1}{\hbar^{2}}\hbar^{3}\rho_{0}^{3}\sum_{j=1}^{N}\sum_{\gamma\neq\delta=1}^{M}\sum_{\alpha'\beta'=1}^{M}\lambda_{\beta'\beta}\lambda_{\alpha'\alpha}\lambda_{\gamma\delta}F_{\alpha\alpha'\beta\beta'\gamma\delta}\delta_{ij} \frac{1}{\omega}\Im I(\omega)\\
= & - \frac{e^{2}}{h}2\pi^{2}\rho_{0}^{3}[(\lambda^{2})_{\alpha\beta}\lambda_{\alpha\beta}-(\lambda^{3})_{\alpha\beta}\delta_{\alpha\beta}]\ln\frac{\hbar\omega}{D_0}\,,
\end{align}
where we took the limit of low frequency $\omega$ and used 
\begin{equation}
\sum_{\gamma\neq\delta=1}^{M}\sum_{\alpha'\beta'=1}^{M}\lambda_{\beta'\beta}\lambda_{\alpha'\alpha}\lambda_{\gamma\delta}F_{\alpha\alpha'\beta\beta'\gamma\delta} =2[(\lambda^{2})_{\alpha\beta}\lambda_{\alpha\beta}-(\lambda^{3})_{\alpha\beta}\delta_{\alpha\beta}]\,.
\end{equation}
In the isotropic case $\lambda_{\alpha\beta}=\lambda(1-\delta_{\alpha\beta})$,
we get $(\lambda^{2})_{\alpha\beta}\lambda_{\alpha\beta}=(M-2)\lambda^{3}(1-\delta_{\alpha\beta})$
and $(\lambda^{3})_{\alpha\beta}\delta_{\alpha\beta}=(M-2)(M-1)\lambda^{3}\delta_{\alpha\beta}$. Finally, 
we obtain the low-frequency conductance to third order in coupling constant, 
\begin{equation}
G_{\alpha\neq\beta}^{ii}(\omega) / G_0=
(\pi \rho_{0} \lambda)^{2} \qty[1 -2 (M-2) \rho_{0} \lambda \ln\frac{\hbar\omega}{D_0}  ]. \label{seq:G3rdOrder}
\end{equation}
We thus find that the leading correction to the conductance diverges in the dc limit, $\omega \to 0$. This divergence is a result of the Kondo renormalization of the coupling constant $\lambda$, described by the first term in the RG equation~(\ref{eq:beta3}) of the main text. 
While the above calculation was done at $T=0$, we expect that at finite temperature, the logarithmic correction would become $\sim \ln \max \{T, \omega \}$. 
This  suggests that in the dc limit the temperature-dependent conductance can be written as $G_{\alpha\neq\beta}^{ii} (T) / G_0=
[\pi  \rho_{0} \lambda(T) ]^{2}$ where we use the renormalized coupling $\lambda (T) \approx \lambda_0 -  (M-2) \rho_0 \lambda_0^2 \ln (T/D_0) $ with cutoff scale $T$. 
Upon lowering the temperature, the correction starts to increase. However, in the large-$N$ limit this increase will be cutoff by higher-order correction to the conductance, which  are captured by the full RG equation~(\ref{eq:beta3}) [or Eq.~(\ref{eq:RG_isotropic})]. The conductance therefore stays small for the entire cross over to $T\to 0$ at which $\lambda(T) \to \pi/(2N)$, where we expect the value $G_{\alpha\neq\beta}^{ii} (0) / G_0=
\pi^{2}/(4N^{2})$. 


\section{Fixed point conductance of multichannel $SO(4)$ topological Kondo model} 
In this section, we derive the fixed point ($T=0$) conductance of the $N$-channel $SO(4)$ topological Kondo model by using a mapping to the 
$2N$CK $SU(2)$ Kondo model, whose conductance was evaluated in Ref.~\cite{PhysRevB.65.195101}. 
Inspired by the mathematical isomorphism $SO(4)\sim SU(2)\times SU(2)$, we will first map the Hamiltonian of 2CK to single channel $SO(4)$ topological Kondo Hamiltonian. 
The 2CK Hamiltonian is
\begin{equation}
H_{\text{2CK}}=J_{\text{2CK}}\Psi^{\text{2CK}\dagger}(s_{x}^{\text{2CK}}S_{x}+s_{y}^{\text{2CK}}S_{y}+s_{z}^{\text{2CK}}S_{z})\Psi^{\text{2CK}},
\end{equation}
where $S_{x},S_{y},S_{z}$ are impurity spin components, $\Psi^{\text{2CK}}=(\psi_{1\uparrow},\psi_{1\downarrow},\psi_{2\uparrow},\psi_{2\downarrow})^{T}$
and 
\begin{equation}
s_{x}^{\text{2CK}}=\left(\begin{array}{cccc}
0 & 1 & 0 & 0\\
1 & 0 & 0 & 0\\
0 & 0 & 0 & 1\\
0 & 0 & 1 & 0
\end{array}\right)\,,\quad s_{y}^{\text{2CK}}=\mi\left(\begin{array}{cccc}
0 & -1 & 0 & 0\\
1 & 0 & 0 & 0\\
0 & 0 & 0 & -1\\
0 & 0 & 1 & 0
\end{array}\right)\,,\quad s_{z}^{\text{2CK}}=\left(\begin{array}{cccc}
1 & 0 & 0 & 0\\
0 & -1 & 0 & 0\\
0 & 0 & 1 & 0\\
0 & 0 & 0 & -1
\end{array}\right).
\end{equation}
Let us map 2CK to 1-channel $SO(4)$ topo Kondo by defining
\begin{equation}
\left(\begin{array}{c}
\psi_{1}\\
\psi_{2}\\
\psi_{3}\\
\psi_{4}
\end{array}\right)=\frac{1}{\sqrt{2}}\left(\begin{array}{cccc}
0 & 1 & -\mi & 0\\
1 & 0 & 0 & \mi\\
\mi & 0 & 0 & 1\\
0 & -\mi & 1 & 0
\end{array}\right)\left(\begin{array}{c}
\psi_{1\uparrow}\\
\psi_{1\downarrow}\\
\psi_{2\uparrow}\\
\psi_{2\downarrow}
\end{array}\right),\ \text{or}\ \Psi^{\text{SO(4)}}=U\Psi^{\text{2CK}},
\end{equation}
with $\Psi^{\text{SO(4)}}=(\psi_{1},\psi_{2},\psi_{3},\psi_{4})^{T}$. 
Then,
\begin{equation}
Us_{x}^{\text{2CK}}U^{\dagger}=\mi\left(\begin{array}{cccc}
0 & 0 & -1 & 0\\
0 & 0 & 0 & 1\\
1 & 0 & 0 & 0\\
0 & -1 & 0 & 0
\end{array}\right),\ Us_{y}^{\text{2CK}}U^{\dagger}=\mi\left(\begin{array}{cccc}
0 & 1 & 0 & 0\\
-1 & 0 & 0 & 0\\
0 & 0 & 0 & 1\\
0 & 0 & -1 & 0
\end{array}\right),\ Us_{z}^{\text{2CK}}U^{\dagger}=\mi\left(\begin{array}{cccc}
0 & 0 & 0 & -1\\
0 & 0 & -1 & 0\\
0 & 1 & 0 & 0\\
1 & 0 & 0 & 0
\end{array}\right).
\end{equation}
The 2CK Hamiltonian becomes
\begin{align}
 & H_{\text{2CK}}/J_{\text{2CK}}=\Psi^{\text{SO(4)}\dagger}U(s_{x}^{\text{2CK}}S_{x}+s_{y}^{\text{2CK}}S_{y}+s_{z}^{\text{2CK}}S_{z})U^{\dagger}\Psi^{\text{SO(4)}}\\
= & \mi(-\psi_{1}^{\dagger}\psi_{3}+\psi_{2}^{\dagger}\psi_{4}+\psi_{3}^{\dagger}\psi_{1}-\psi_{4}^{\dagger}\psi_{2})S_{x}+\mi(\psi_{1}^{\dagger}\psi_{2}+\psi_{3}^{\dagger}\psi_{4}-\psi_{2}^{\dagger}\psi_{1}-\psi_{4}^{\dagger}\psi_{3})S_{y}+\mi(-\psi_{1}^{\dagger}\psi_{4}-\psi_{2}^{\dagger}\psi_{3}+\psi_{3}^{\dagger}\psi_{2}+\psi_{4}^{\dagger}\psi_{1})S_{z}.
\end{align}
Next, we represent the spin operators by four Majoranas with a fixed total fermion parity: 
 $\gamma_{1}\gamma_{2}\gamma_{3}\gamma_{4}=1$
or $\gamma_{4}=\gamma_{1}\gamma_{2}\gamma_{3}$. 
\begin{equation}
S_{x}=\frac{\mi}{2}\gamma_{2}\gamma_{4}=-\frac{\mi}{2}\gamma_{1}\gamma_{3},\ S_{y}=\frac{\mi}{2}\gamma_{3}\gamma_{4}=\frac{\mi}{2}\gamma_{1}\gamma_{2},\ S_{z}=-\frac{\mi}{2}\gamma_{1}\gamma_{4}=-\frac{\mi}{2}\gamma_{2}\gamma_{3}.
\end{equation}
We can check that 
\begin{align}
[S_{x},S_{y}]= & \frac{1}{4}(\gamma_{1}\gamma_{3}\gamma_{1}\gamma_{2}-\gamma_{1}\gamma_{2}\gamma_{1}\gamma_{3})=\frac{1}{2}\gamma_{2}\gamma_{3}=\mi S_{z},\\{}
[S_{x},S_{z}]= & -\frac{1}{4}(\gamma_{1}\gamma_{3}\gamma_{2}\gamma_{3}-\gamma_{2}\gamma_{3}\gamma_{1}\gamma_{3})=\frac{1}{2}\gamma_{1}\gamma_{2}=-\mi S_{y},\\{}
[S_{y},S_{z}]= & \frac{1}{4}(\gamma_{1}\gamma_{2}\gamma_{2}\gamma_{3}-\gamma_{2}\gamma_{3}\gamma_{1}\gamma_{2})=\frac{1}{2}\gamma_{1}\gamma_{3}=\mi S_{x}
\end{align}
which means that they satisfy the $SU(2)$ algebra as they should. 
By expanding 2CK Hamiltonian
\begin{align}
H_{\text{2CK}}/J_{\text{2CK}}= & -\frac{1}{2}\gamma_{1}\gamma_{3}\psi_{1}^{\dagger}\psi_{3}-\frac{1}{2}\gamma_{2}\gamma_{4}\psi_{2}^{\dagger}\psi_{4}-\frac{1}{2}\gamma_{3}\gamma_{1}\psi_{3}^{\dagger}\psi_{1}-\frac{1}{2}\gamma_{4}\gamma_{2}\psi_{4}^{\dagger}\psi_{2}\\
 & -\frac{1}{2}\gamma_{1}\gamma_{2}\psi_{1}^{\dagger}\psi_{2}-\frac{1}{2}\gamma_{3}\gamma_{4}\psi_{3}^{\dagger}\psi_{4}-\frac{1}{2}\gamma_{2}\gamma_{1}\psi_{2}^{\dagger}\psi_{1}-\frac{1}{2}\gamma_{4}\gamma_{3}\psi_{4}^{\dagger}\psi_{3}\\
 & -\frac{1}{2}\gamma_{1}\gamma_{4}\psi_{1}^{\dagger}\psi_{4}-\frac{1}{2}\gamma_{2}\gamma_{3}\psi_{2}^{\dagger}\psi_{3}-\frac{1}{2}\gamma_{3}\gamma_{2}\psi_{3}^{\dagger}\psi_{2}-\frac{1}{2}\gamma_{4}\gamma_{1}\psi_{4}^{\dagger}\psi_{1}\\
= & -\sum_{i\neq j}^{4}\gamma_{i}\gamma_{j}\psi_{i}^{\dagger}\psi_{j},
\end{align}
we see the equivalence to the 1-channel $SO(4)$ topological Kondo Hamiltonian. 
The mapping is straightforwardly generalized to show the equivalence between $N$-channel $SO(4)$ topological Kondo and $2N$CK models.

Next, we connect the  fixed point conductances of $2N$CK and $N$-channel $SO(4)$ topological Kondo models by using the CFT method~\cite{PhysRevLett.109.156803,Oshikawa_2006}. 
We seek the general result beyond the weak tunneling limit. Therefore, we introduce the left- and right-moving currents at position $x$ as  $I_{Xj}(x)=\psi_{Xj}^{\dagger}(x)\psi_{Xj}(x)$ where $X=L,R$ labels the left/right movers in the linearized spectrum of the (flavor) ``lead'' $j=1,\dots,4$, in channel 1.  
(Even though we consider $N$-channel $SO(4)$ topological Kondo, the conductance in the main text is measured between different flavors in the first channel.)  
The conductance matrix can then be evaluated by using the Kubo formula, 
\begin{equation}
G_{jj'}=\Re\lim_{\omega\to0}\frac{1}{\hbar\omega}\int_{0}^{\infty}\mathrm{d}t\,\mathrm{e}^{\mathrm{i}\omega t}\frac{1}{L}\int_{0}^{L}\mathrm{d}x \langle[I_{j}(y,t),I_{j'}(x,0)]\rangle\,, \quad I_{j} = I_{Rj} - I_{Lj}\,, \label{seq:Gjj}
\end{equation}
where voltage is applied to a region $0<x<L$ in the lead $j'$ and $y$ is the point in lead $j$ where the current is measured (but whose value does not matter)~\cite{Oshikawa_2006}. 
 The current correlation functions can be evaluated in the free case in the absence of a Kondo impurity. 
In that case, the conductance matrix vanishes as no current is carried, $G_{jj}^{\text{free}} = G_{j\neq j'}^{\text{free}} =  0$. It is nevertheless convenient to introduce the following ``L/R conductance'' that characterizes the backscattering of left-moving currents into right-moving ones at the end of the wire (from hereon, we focus on $j=1$ for simplicity),  
\begin{equation}
G_{L,R;11}^{\text{free}}=  \Re\lim_{\omega\to0}\frac{1}{\hbar\omega L}\int_{0}^{L}\mathrm{d}x\int_{0}^{\infty}\mathrm{d}t\,\mathrm{e}^{\mathrm{i}\omega t}\langle[I_{R,1}(y,t),I_{L,1}(x,0)]\rangle_{\text{free}} = \frac{e^2}{h}\,,
\end{equation}
where we used the correlation function ($v_F$ is the Fermi velocity)
\begin{equation}
 \langle[I_{R,1}(y,t),I_{L,1}(x,0)]\rangle_{\text{free}} = \mathrm{i} \frac{e^{2}v_{F}^{2}}{(2\pi)^{2}}  \frac{4\eta(y+x-v_{F}t)}{[\eta^{2}+(y+x-v_{F}t)^{2}]^{2}}\,. \label{seq:IIfree}
\end{equation}
Here, $\langle \cdots \rangle_{\text{free}}$ is the zero-temperature average over the free Hamiltonian which contains only kinetic terms and no Kondo interaction. In addition, a boundary condition was applied that connects left and right mover so that their correlation function above does not vanish \cite{Oshikawa_2006}.

We wish to evaluate the conductance, Eq.~(\ref{seq:Gjj}) with $j\neq j'$, in the Kondo intermediate coupling fixed point.  Due to the $SO(M)$ symmetry of the flavors, we only need to calculate  $G_{12}^{\text{Kondo}}$.
Furthermore,  only the $LR$ current correlation function contributes to the off-diagonal conductance, so $G_{12}^{\text{Kondo}} = G_{L,R;12}^{\text{Kondo}}$. 
As pointed out for example in Refs.~\cite{PhysRevLett.109.156803,PhysRevLett.67.3160,affleck1995conformal,PhysRevB.65.195101}, the correlation functions at the Kondo intermediate coupling fixed point are proportional to their free fixed point values, Eq.~(\ref{seq:IIfree}), up to a constant factor, given by the modular S-matrix. 
For example, we show below that 
\begin{equation}
    \langle I_{L1}I_{R2}\rangle_{\text{Kondo}} = 
 \frac{1-S_{N}}{4} \langle I_{L1}I_{R1}\rangle_{\text{free}} \,, \label{seq:IKIfratio}
\end{equation} 
where $S_N=\cos[3\pi/(2N+2)]/\cos[\pi/(2N+2)]$ is given by the appropriate elements of the modular S-matrix. 
The arguments of $\langle I_{L,i}(y,t)I_{R,j}(x,0)\rangle$ are hidden for brevity, as we mostly do not need them. The correlator $\langle \cdots \rangle_{\text{Kondo}}$ denotes the average over the  Hamiltonian with Kondo interaction at the Kondo coupling fixed point. 
This indicates that the desired conductance can be obtained by considering the ratio of the corresponding current correlation function and $\langle I_{L1}I_{R2}\rangle_{\text{free}}$:  
\begin{equation}
    \frac{G_{12}^{\text{Kondo}}}{G_{L,R;11}^{\text{free}}}=
    \frac{ \langle I_{L1}I_{R2}\rangle_{\text{Kondo}} }{\langle I_{L1}I_{R1}\rangle_{\text{free}}}=\frac{1-S_N}{4}=\sin^2 \frac{\pi}{2N+2}, \label{seq:so4conduc}
\end{equation}
which gives Eq.~(\ref{eq:Nso4}) of the main text. 

\subsection{Derivation of the relation~(\ref{seq:IKIfratio})}

In order to derive Eq.~(\ref{seq:IKIfratio}), we follow Ref.~\cite{PhysRevLett.109.156803} and introduce the 
densities $J_{X}^{(c)}(x)=\Psi_{X}^{\text{SO(4)}\dagger}(x)\Psi_{X}^{\text{SO(4)}}(x)$,
$J_{X}^{(d)}(x)=\Psi_{X}^{\text{SO(4)}\dagger}(x)D_{1}\Psi_{X}^{\text{SO(4)}}(x)$,
 and $D_{1}=\text{diag}(1,-1,-1,1)$ for the  $SO(4)$ model. 
 The  correlation functions of charge currents can be expressed in terms of these densities, see Eq.~(\ref{seq:currentmapping}) below. 
 Moreover, these densities can be expressed in the 2CK basis as $J_{X}^{(c)}=\Psi_{X}^{\text{2CK}\dagger}\Psi_{X}^{\text{2CK}}$ and 
 $J_{X}^{(d)}=-\Psi_{X}^{\text{2CK}\dagger}D_{1}\Psi_{X}^{\text{2CK}}$. Therefore, we can use the well-known \cite{affleck1995conformal} relations between correlation functions of the Kondo (K) and free (f) fixed points in the MCK model. 
 We give these results in the  $2N$-channel case which can be mapped to $N$-channel $SO(4)$ topological Kondo model. 
 The operator $J_{X}^{(c)}$ transforms like an $SU(2)$ singlet (it does not couple to the impurity) and therefore satisfies 
 $\langle J_{L}^{(c)}J_{R}^{(c)}\rangle_{\text{K}}=\langle J_{L}^{(c)}J_{R}^{(c)}\rangle_{\text{f}}$, 
 while $J_{X}^{(d)}$ transforms in the adjoint (spin-1) representation and thus 
 $\langle J_{L}^{(d)}J_{R}^{(d)}\rangle_{\text{K}}=S_{N}\langle J_{L}^{(d)}J_{R}^{(d)}\rangle_{\text{f}}$ where $S_N = S^1_{1/2} S^0_{0} / (S^1_{0} S^0_{1/2} )  =\cos[3\pi/(2N+2)]/\cos[\pi/(2N+2)]$ is given by elements of the modular S-matrix for 2$N$CK~\cite{PhysRevB.65.195101}.

Now we only need to express the charge current correlation functions in terms of the above correlations functions.  
 In the isotropic case, we 
have $\langle I_{Lj}I_{Rj}\rangle=\langle I_{L1}I_{R1}\rangle$
for all $j$ and $\langle I_{Lj}I_{Rj'}\rangle=\langle I_{L1}I_{R2}\rangle$
for all $j\neq j'$. By these definitions and constrains for the current correlation functions, we find 
\begin{equation}
\langle J_{L}^{(c)}J_{R}^{(c)}\rangle=  4\langle I_{L1}I_{R1}\rangle+12\langle I_{L1}I_{R2}\rangle,\ \langle J_{L}^{(d)}J_{R}^{(d)}\rangle=4\langle I_{L1}I_{R1}\rangle-4\langle I_{L1}I_{R2}\rangle. \label{seq:currentmapping}
\end{equation}
At the free case, $\langle I_{L1}I_{R2}\rangle_{\text{f}}=0$ which simplifies Eq.~(\ref{seq:currentmapping}) to 
\begin{equation}
    \langle J_{L}^{(c)}J_{R}^{(c)}\rangle_{\text{f}}= 4\langle I_{L1}I_{R1}\rangle_{\text{f}},\ \langle J_{L}^{(d)}J_{R}^{(d)}\rangle_{\text{f}}=4\langle I_{L1}I_{R1}\rangle_{\text{f}}\,.
\end{equation}
At the Kondo fixed point, by inverting Eq.~(\ref{seq:currentmapping}) 
we get
\begin{align}
& \langle I_{L1}I_{R1}\rangle_{\text{K}}=\frac{1}{16}\langle J_{L}^{(c)}J_{R}^{(c)}\rangle_{\text{K}} + \frac{3}{16} \langle J_{L}^{(d)}J_{R}^{(d)}\rangle_{\text{K}}=\frac{1+3 S_N}{4} \langle I_{L1}I_{R1}\rangle_{\text{f}}\,, \\
& \langle I_{L1}I_{R2}\rangle_\text{K}=\frac{1}{16}\langle J_{L}^{(c)}J_{R}^{(c)}\rangle_{\text{K}} - \frac{1}{16} \langle J_{L}^{(d)}J_{R}^{(d)}\rangle_{\text{K}}=\frac{1-S_N}{4} \langle I_{L1}I_{R1}\rangle_{\text{f}}\,.
\end{align}
Thus, we have derived Eq.~(\ref{seq:IKIfratio}).

\section{Impurity entropy from CFT}
The impurity entropy is given as $S_{\text{imp}}=\ln g$. Kimura \cite{doi:10.7566/JPSJ.90.024708,doi:10.7566/JPSJ.86.084703} has calculated $g$ for general $N$-channel $SO(M)$ model (denoted spinor $SO(M)_{2N}$ in Ref.~\cite{doi:10.7566/JPSJ.90.024708}). 
They used one of the main results from CFT~ \cite{doi:10.7566/JPSJ.90.024708,doi:10.7566/JPSJ.86.084703,affleck1995conformal}, which is $g=S^0_s/S^0_0$ where $S^j_s$ is the modular S-matrix. The modular S-matrix of $SO(M)$ is given by Hung et al \cite{hung2018linking} and the general information of it can also be found \cite{francesco2012conformal}. Here, $s$ means the representation for the impurity. In terms of Dynkin labels, we can take $s=(0,\cdots,0,1)$ for both of odd and even $M$ cases. Thus, we get $g$ for arbitrary $M,N$:
\begin{equation}
g(M,N)=\begin{cases}
2^{p}\prod_{i=0}^{p-1}\cos\Big[\frac{\pi(i+1/2)}{2N+2p-1}\Big], & M=2p+1\,\text{is odd}\\
2^{q-1}\prod_{i=1}^{q-1}\cos\Big[\frac{\pi i}{2N+2q-2}\Big], & M=2q\,\text{is even}
\end{cases}
\end{equation}

The first thing we can check is the large-$N$ limit where we can consider 
\begin{align}
g(2p+1,N)= & 2^{p}\prod_{i=0}^{p-1}\cos\Big[\frac{\pi(i+1/2)}{2N+2p-1}\Big]\approx2^{p}\prod_{i=0}^{p-1}\qty[1-\frac{\pi^{2}(i+1/2)^{2}}{8N^{2}}]\\
\approx & 2^{p}\Big[1-\sum_{i=0}^{p-1}\frac{\pi^{2}(i+1/2)^{2}}{8N^{2}}\Big]=2^{p}\Big[1-\frac{p(2p-1)(2p+1)\pi^{2}}{96N^{2}}\Big].\\
g(2q,N)= & 2^{q-1}\prod_{i=1}^{q-1}\cos\Big(\frac{\pi i}{2N+2q-2}\Big)\approx2^{q-1}\prod_{i=1}^{q-1}\Big(1-\frac{\pi^{2}i^{2}}{8N^{2}}\Big)\\
\approx & 2^{q-1}\Big(1-\sum_{i=1}^{q-1}\frac{\pi^{2}i^{2}}{8N^{2}}\Big)=2^{q-1}\Big[1-\frac{q(q-1)(2q-1)\pi^{2}}{48N^{2}}\Big].
\end{align}
These two equations give Eq.~(\ref{eq:largeNimpurity}) in the main text. When taking $N\to\infty$, $g(2p+1,N)=2^{p}$
is the dimension of the spinor representation of $SO(2p+1)$ and $g(2q,N)=2^{q-1}$
is also the dimension of the spinor representation (two spinor representations have the same dimension), which means that the island is isolated from the itinerant electrons so that the dimension is totally given by the quantum dimension of Majoranas.

Let us take a look at $N=1$ and $N=2$
case. The first one is mentioned by Altland et al.~\cite{Altland_2014} that
\begin{equation}
g(M,1)=\begin{cases}
\sqrt{2p+1}=\sqrt{M}, & M=2p+1\,\text{is odd}\\
\sqrt{q}=\sqrt{M/2}, & M=2q\,\text{is even}
\end{cases}
\end{equation}
Based on this, we can calculate the second case by using
\begin{equation}
g(2p+1,1)=2^{p}\prod_{i=0}^{p-1}\cos\Big[\frac{\pi(i+1/2)}{2p+1}\Big]=\sqrt{2p+1},\ \text{and}\ g(2q,1)=2^{q-1}\prod_{i=1}^{q-1}\cos\Big(\frac{\pi i}{2q}\Big)=\sqrt{q}.
\end{equation}
Thus, when $N=2$, we have
\begin{align}
g(2p+1,2) & =2^{p}\prod_{i=0}^{p-1}\cos\Big[\frac{\pi(i+1/2)}{2p+3}\Big]=\frac{g(2p+3,1)}{2\cos\Big[\frac{\pi(2p+1)}{2(2p+3)}\Big]}=\frac{\sqrt{2p+3}}{2\cos\Big[\frac{\pi(2p+1)}{2(2p+3)}\Big]}.\\
g(2q,2) & =2^{q-1}\prod_{i=1}^{q-1}\cos\Big[\frac{\pi i}{2q+2}\Big]=\frac{g(2q+2,1)}{2\cos\Big[\frac{\pi q}{2(q+1)}\Big]}=\frac{\sqrt{q+1}}{2\cos\Big[\frac{\pi q}{2(q+1)}\Big]}.
\end{align}
In terms of $M$, they are
\begin{equation}
g(M,2)=\begin{cases}
\frac{1}{2}\sqrt{M+2}/\cos\big[\frac{\pi M}{2(M+2)}\big], & M\,\text{is odd}\\
\frac{1}{2}\sqrt{(M+2)/2}/\cos\big[\frac{\pi M}{2(M+2)}\big], & M\,\text{is even}
\end{cases}
\end{equation}
which is Eq.~(\ref{eq:gN2}) in the main text. As we know, $N=1$ spinor $SO(4)$ is like 2-channel Kondo. It is natural to think $N=2$ spinor $SO(4)$ as 4-channel Kondo. They indeed have the same $g$, which is $\sqrt{3}\approx1.732$. Also, as Beri et al. \cite{PhysRevLett.109.156803} mentioned that $N=1$ spinor $SO(3)$ maps to 4-channel Kondo, we verify that $N=2$ spinor $SO(3)$ maps to 8-channel Kondo which has $g(3,2)=\sqrt{(5+\sqrt{5})/2}=1.90211$. 


\section{More details on flavor and channel anisotropies}
In this Section, we give more details on the flavor anisotropy (Sec.~\ref{SMsec:flavor}) and the channel anisotropy  in the case of $\mathcal{O}(1)$ anisotropic channels (Sec.~\ref{SMsec:N2}). 

\subsection{Tunneling anisotropy induced flavor anisotropy \label{SMsec:flavor}}
In the main text, we considered a specific kind of flavor anisotropy and showed that it is irrelevant, see the flow diagram in Fig.~\ref{fig:anisotropy}a. 
Here, we explain the physical consideration behind using that exact anisotropy and derive the RG equations that were used to obtain the flow diagram.

In our model, not any anisotropy is physically realistic due to the form of  $\lambda_{i,\alpha\beta}$ in Eq.~(\ref{eq:V}) [see also Eq.~(\ref{eq:V2C})], defined by the product $t^i_{\alpha}t^{i}_{\beta}$ of real tunneling amplitudes. 
It means that the corresponding anisotropy should originate from the anisotropy of $t_{\alpha} =(\vec{t})_{\alpha} $ instead of from $\lambda_{\alpha\beta}$ directly. Here we consider the channel-isotropic case and therefore suppress the channel index $i$ in $t^i_{\alpha}$ and  $\lambda_{i,\alpha\beta}$.
For all $N$ channels, let us consider anisotropic tunneling $\vec{t}=(t',t,\cdots,t)$, corresponding to a different coupling to flavor $\alpha = 1$. 
This form leads to the flavor anisotropy presented in the main text,  $\lambda_{\alpha\beta}=[\lambda+(\lambda'-\lambda)(\delta_{1\alpha}+\delta_{1\beta})]$. 
 We find from Eq.~(\ref{eq:beta3}) of the main text, 
\begin{align}
\dv{\lambda}{l}=(1-2N\lambda\rho_0)\qty[(M-3)\lambda^2+\lambda'^2]\rho_0,\ \
\dv{\lambda'}{l}=(M-2)\lambda'\rho_0\qty(\lambda-N\lambda^2\rho_0-N\lambda'^2\rho_0).
\end{align}
From  these  RG equations, we find  two nontrivial fixed points with the majority coupling $\lambda =1/(2N\rho_0)$ in both, while the minority coupling is either $\lambda' =1/(2N \rho_0)$ or $\lambda' =0$ (see Fig. \ref{fig:anisotropy}a of the main text). 
The first one is isotropic and stable, while the    second fixed point is unstable. 
This shows that flavor anisotropy remains irrelevant  in the multichannel generalization of the topological Kondo model.

\subsection{Channel anisotropy when $N_2$ is not large \label{SMsec:N2}}

In the main text, we discussed the channel anisotropy with $N_1,N_2$ both large. In this section, we only consider $N_1$ large but $N_2$ of order $1$. 
We then obtain from Eq.~(\ref{eq:channelaniso}) of the main text, \begin{equation}
\dv{\lambda_1}{l}  = (M-2)\lambda_1^2 \rho_0(1-2N_1\lambda_1 \rho_0)\,, \quad  
\dv{\lambda_2}{l}  = (M-2)\lambda_2^2 \rho_0(1-2N_1 \frac{\lambda_1^2}{\lambda_2} \rho_0)\,.
\end{equation} 
In the isotropic case,  $\lambda_1 = \lambda_2$, we recover Eq.~(\ref{eq:RG_isotropic}) of the main text. 
When $\lambda_1>\lambda_2$, the  isotropic fixed point is unstable and the flow is towards the fixed point  $\lambda_{1}=1/(2N_1),\lambda_{2}=0$. When $\lambda_1<\lambda_2$, the coupling $\lambda_2$ flows towards  strong coupling. 
The corresponding RG flow diagram is plotted in Fig.~\ref{fig:channelN2}.
\begin{figure}[tb]
\includegraphics[width=0.45\columnwidth]{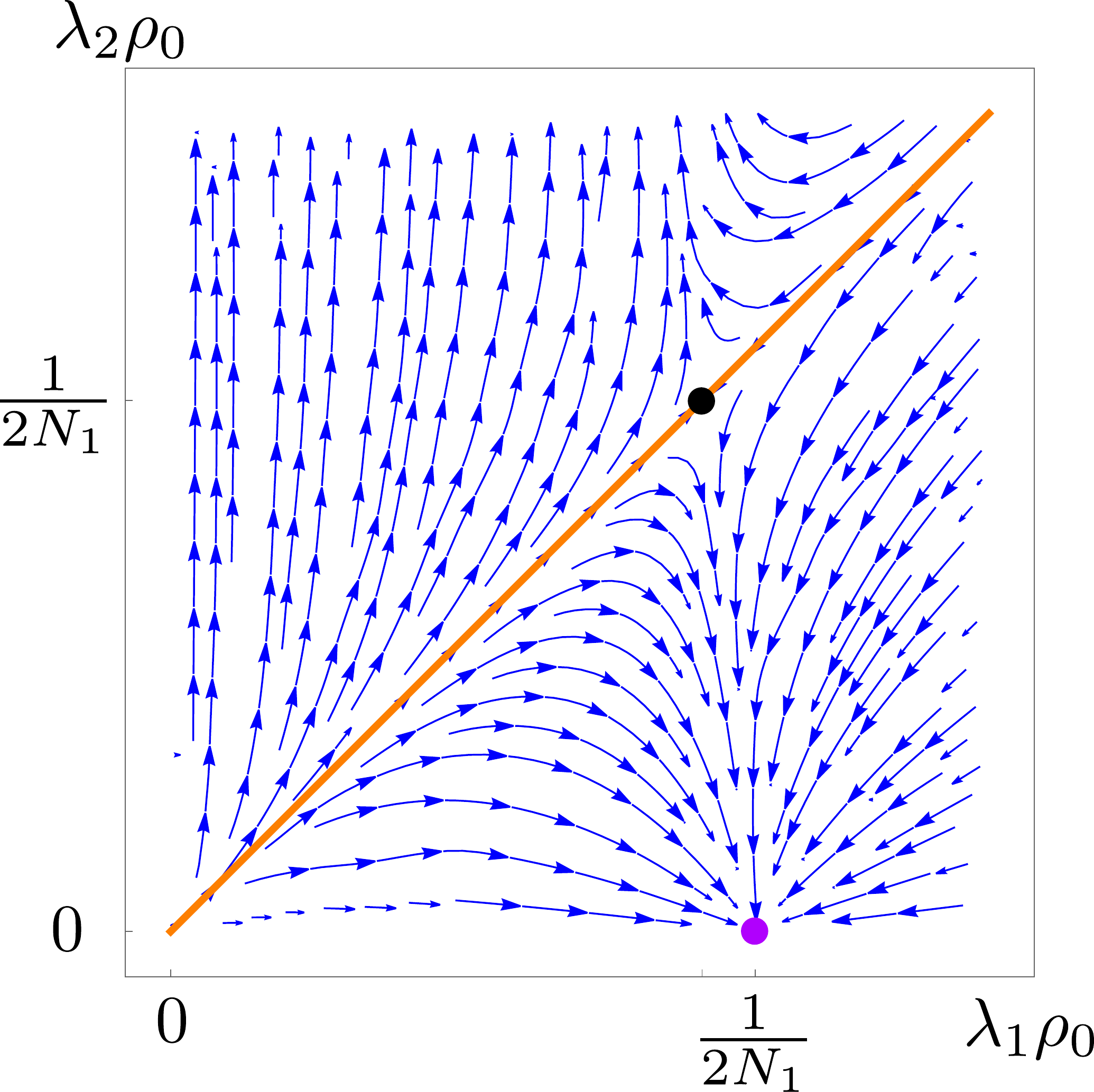}
\caption{RG flow diagram of channel anisotropy in the large-$N_1$ limit with  $N_2\sim \mathcal{O}(1)$. The isotropic fixed point (black point) is at $\lambda_1=\lambda_2 =1/2[(N_1+N_2)\rho_0]$ and is  unstable. When $\lambda_1>\lambda_2$,  the stable fixed point is the purple one [$\lambda_1=1/(2N_1\rho_0),\lambda_2=0$] while when $\lambda_1<\lambda_2$, the coupling $\lambda_2$ flows to strong coupling.} 
\label{fig:channelN2}
\end{figure}



\end{widetext}

\end{document}